# Synchrotron radiation contributions to optical diffraction radiation measurements

*G.A. Naumenko[\*].*

Institute for Nuclear Physics, pr.Lenina 2a, 634050, Tomsk, Russia.

___________________________________________________________________

**Abstract**

If we try to measure the backward optical diffraction radiation (BODR) of high energy electrons from a conductive slit or a semi infinitive plate, the electron beam will pass thru the bending or steering magnets or magnet lenses before striking the target. The synchrotron radiation (SR) from these magnets can obscure the BODR measurements. An analysis of the properties of SR from these magnets is in this paper presented. A model based on the modified Lenar Wikherd potentials was created, and the SR angular distribution from relativistic electrons in bending and steering magnets for different conditions of radiation in the optical region was calculated. The analysis shows, that for the conditions of the KEK ATF extraction line, the intensity of SR exceeds that of the backward optical transition radiation (BOTR) from the conducting targets, and it is much lager than the intensity of the BODR. The SR intensity from the steering magnets depends on its tuning and may be comparable to BOTR. Thus, these results it is seen, that the problem of separation of the BODR and SR in the BODR measurements is important. Two methods resolving of this problem is in this article suggested.

---

[\*] Corresponding author E-mail: naumenko@npi.tpu.ru

PACS: 41.60-m, 42.79 Dj

Keywords: Relativistic electrons, synchrotron radiation, diffraction radiation

## 1. Introduction

Recently, attention has been devoted to using BODR as a new non-invasive tool for ultrarelativistic beams. Castellano [1] considered the possibility of determining the transverse beam size using BODR characteristics when the beam moves through a slit in a perpendicular screen. It was suggested that the BODR angular distribution be measured in the vicinity of the initial electron beam direction. Fiorito et al.[3] have suggested that the BODR generated by a particle passing through a circular aperture in a conducting screen be used for the noninvasive diagnostics of the beams position, size, and divergence. Ref [2] considered the possibility of carrying out a proof-of-principle experiment at KEK-ATF for estimating the prospects of BODR usage for non-invasive beam diagnostics. Thus, the BODR monitor is a promising candidate for single-pulse beam-profile measurements, and that it will be an extremely useful instrument for future linear colliders (e.g. JLC, NLC,TESLA and CLIC).

However, synchrotron emission contribution may be obstacle for the realization of these projects. We can see this, if we try to measure the BODR of relativistic electrons of more then 200 MeV energy from conductive slit ore semi infinitive plate. The electron beam passes the bending or steering magnets or magnet lenses before striking the target (see Fig.1).

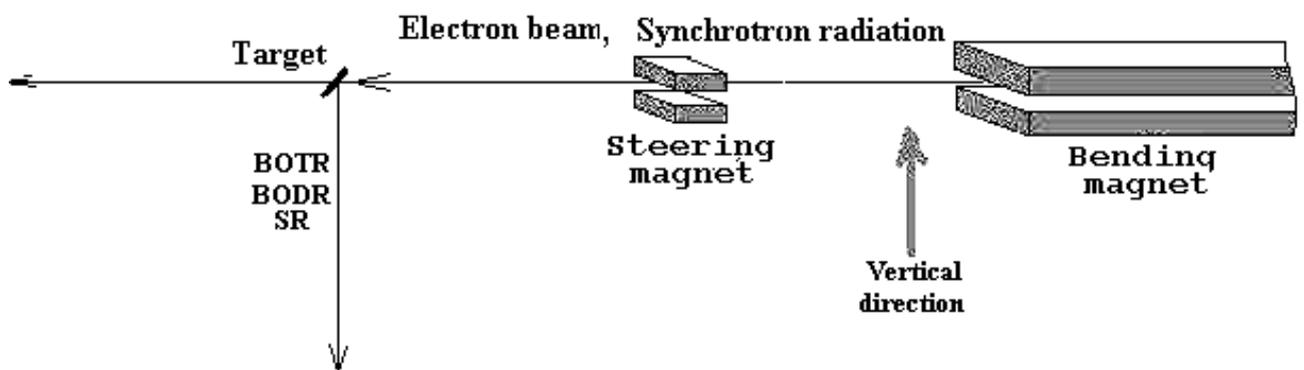

**Fig.1**

*The geometry of SR production for BODR investigations.*

Thus in the target region there will be synchrotron radiation (SR) from these magnets. This radiation may reflect from the target passing in the same direction as the investigated radiation. How intensive this radiation is in comparison with the investigated BODR is important, because it may be obstacle for the BODR measurements. Therefore, the analysis of the properties of SR from these magnets is necessary. Usually one considers the SR from the middle of bending magnets ([4]). In our case SR is generated from the edge of magnets, when electrons transit from magnetic field into free motion. Ref [5] considered the SR emitted at the edge of magnet using simple dipole approach, which can not to be applied to large sources such a bending magnet. The SR characteristics depend on the experimental conditions (e.g. bending radius, electron energy, and length of bending trajectory). Therefore an elaboration of the theoretical model for specified conditions is necessary.

## 2. The theoretical model for the calculation of the properties of SR from the edge of magnets.

For building of the theoretical model let me to use the Fourier presentation of the well known Lenar-Wikerd potentials in relativistic approach (Lorents-factor $\gamma \gg 1$):

$$\vec{E}_w = \frac{e^{-iwR_0}}{4\pi R_0} \cdot \frac{e^-}{(1-\beta(\vec{n}\cdot\vec{v}))^2} \cdot \left[\vec{n}\times\left[(\vec{n}-\beta\vec{v})\times\vec{a}(w')\right]\right], \quad (1)$$

where $w' = w\cdot(1-\beta(\vec{n}\cdot\vec{v}))$, $\vec{v}$ is the vector of electron velocity direction, $\vec{n}$ is the vector of observation direction, $\beta$ is a module of electron velocity, $e^-$ is electron charge, $w$ is the frequency of investigated radiation, and $R_0$ is the distance between radiator and observer. In (1) and further in the paper use is made of the system of units $\hbar=c=1$, unless otherwise indicated. In our analysis we do not use the dipole approach because in our case the size of SR source is comparable to the distance between source and observation point. Therefore we will divide the bending magnet into a series of many small magnets (so, that for each magnet we can use the dipole approach). We will then sum the field of all magnets. In integral limit we obtain:

$$\vec{E}_w = \frac{e^{-iwR_0}}{4\pi R_0} \cdot \frac{e^-}{\sqrt{2\pi}} \cdot \int e^{-iw't} \frac{\left[\vec{n}\times\left[(\vec{n}-\beta\vec{v}(t))\times\vec{a}(t)\right]\right]}{(1-\beta(\vec{n}\cdot\vec{v}(t)))^2} dt, \quad (2)$$

here $w' = w\cdot(1-\beta(\vec{n}\cdot\vec{v}(t)))$

The spectral and angular density distribution of radiation intensity is:

$$\frac{dW}{dw\cdot d\Omega} = 4\pi\cdot R^2 \left|\vec{E}_w\right|^2 \quad (3)$$

We cannot obtain from (3) the full analytical expression for real experimental condition, but it is possible to calculate it numerically. Below we present the numerically calculations of the SR properties for different experimental conditions. For checking of this model we calculate the distribution of SR from the middle of magnet end compare it with the similar calculations using a well-known model [4]. Fig.2 is shows the spectra-angular distribution of SR from the middle of magnet with deflecting radius 8 m for electron energy $E_0 \approx 1.26$ GeV and for the length of magnetic field 1.28 m, calculated using our model.

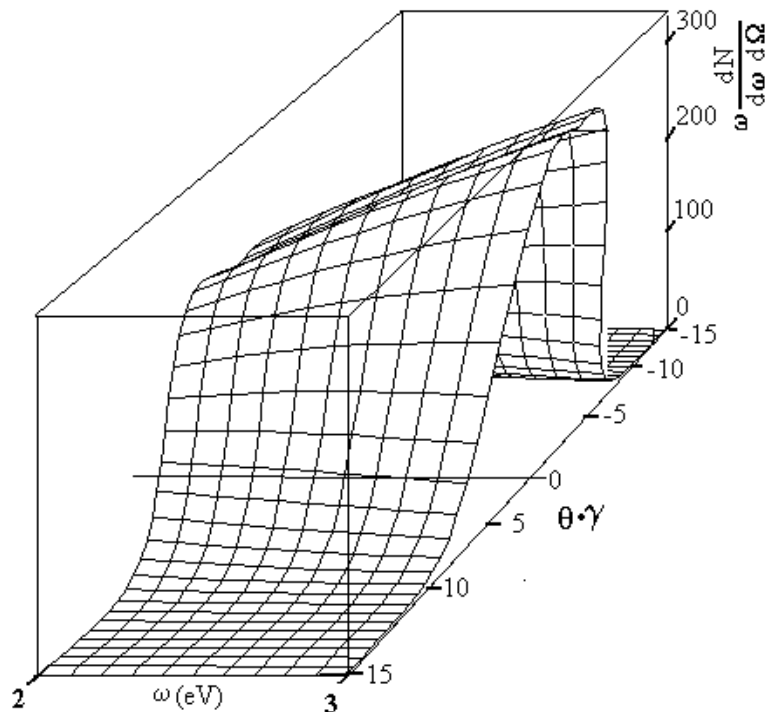

**Fig.2**

*Spectra – angular distribution of SR from the middle of magnet, calculated using (3). è is vertical observing angle*

A similar distribution, calculated using formula from [4], gives the same results.

### 3. Properties of synchrotron radiation from edge of bending magnet for KEK ATF extraction line condition.

Let us consider the example of SR properties for the experimental conditions of the KEK ATF extraction line. Let us also consider the horizontal and vertical component of the SR electromagnetic field (horizontal an vertical SR polarization) in terms of the geometry shown in Fig.1.

The main properties of KEK ATF extraction line, related to our problem are: the energy of accelerated electrons $E_0$=1.26 GeV ($\gamma$=2470), the inflection radius of electron orbit in bending magnet (BM) $R$=8 m, and the length of BM magnetic field along the electron trajectory $L$=1.28 m.

Our first calculations show that if $R/\gamma \ll L$, the SR properties from the edge of magnet for these conditions do not depend on $L$. Also the spectral dependence in the optical region is weak. Thus, for KEK ATF conditions only two factors ($\gamma$ and $R$) influence on the SR properties. Fig.3 shows the SR intensity angular distribution calculated using (3) for above-mentioned conditions.

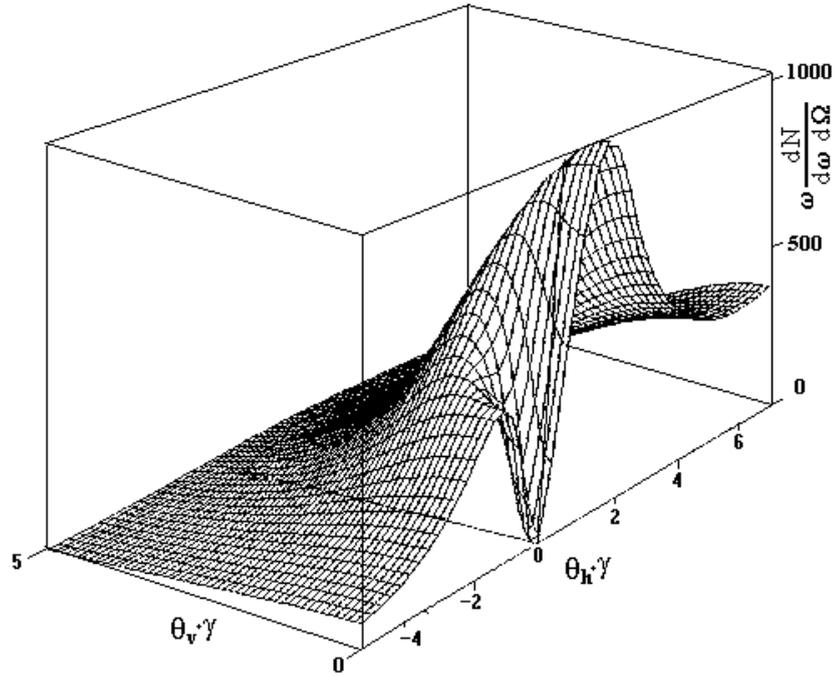

**Fig.3**

*Angular distribution of SR from the edge of bending magnet of KEK ATF extraction line. $q_h$ - horizontal angle, $q_v$ - vertical one (see Fig.1)*

In Fig.3, and further in the paper, the angles are presented in units of $\gamma^{-1}$. Here $\theta_h = 0$ corresponds to the outgoing electron velocity direction and negative $\theta_h$ corresponds to the direction of electron deflection. For the large positive $\theta_h$ the SR angular distribution transforms to the case of SR from the middle of the BM (see Fig.2). The radiation intensity at the angle $\theta_h \approx 1$ exceeds significantly the radiation from the middle of magnet. Fig.4 and Fig.5 shows the angular distribution of SR intensity of both polarizations. We see from these figures, that intensity of horizontal polarization exceeds approximately two times the one of vertical polarization.

These results were obtained for an abrupt edge of the magnetic field, but a defused field on the edge of magnets is more appropriate. Therefore an estimation of possible contribution of the defused part of magnetic field should be made.

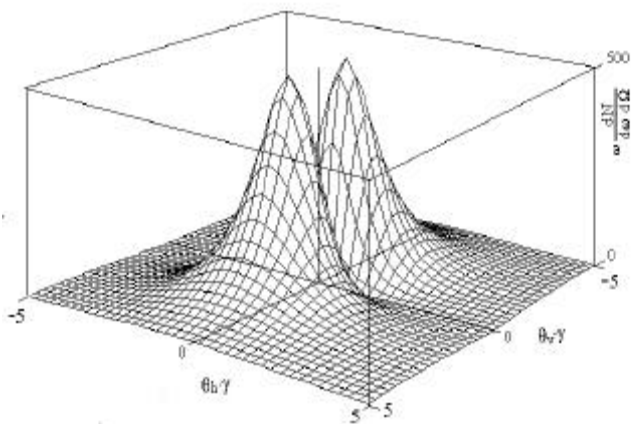

**Fig.4**

*Angular distribution of the vertical polarization of the SR intensity. $q_h$ - horizontal angle, $q_v$ - vertical one (see Fig.1)*

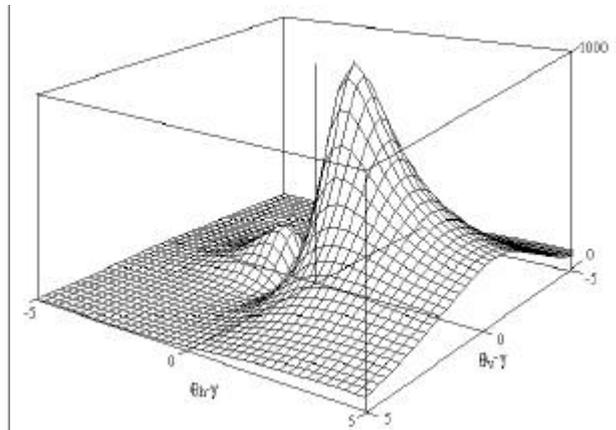

**Fig.5**

*Angular distribution of the horizontal polarization of the SR intensity. $q_h$ - horizontal angle, $q_v$ - vertical one (see Fig.1)*

We considered for this purpose the approach of linear dissipation of defused fields at the edge of magnet. The calculation of SR distribution is illustrated on Fig.3,4,5 shows, that for the size of defused part of magnetic field less than $5R/\gamma$ the deviation of SR distribution does not exceeds 5%.

We will now compare the intensity of possible types of radiation, which may be observed in case of BODR investigation. Fig.6 shows the horizontal angular distribution of intensity of SR, BOTR and BODR.

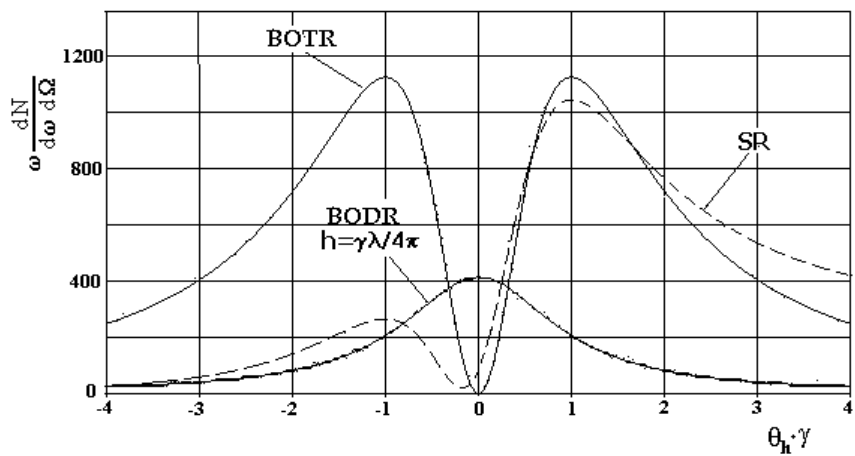

**Fig.6**

*Horizontal angular distribution of the intensity of SR, BOTR and BODR $q_h$ is horizontal angle, vertical angle $q_v=0$*

The BOTR intensity was calculated using [6]. An approach of ideally conducting and infinite thin plane was used [6]. The BODR intensity from semi-infinite plane was calculated using formula from [7] using the same approach for the characteristic impact parameter $h=\gamma\lambda/4\pi$.

These distributions show that the SR differential intensity is comparable to the BOTR and may be much lager than the intensity of the BODR. Thus, the problem of separating BODR and SR for BODR experimental investigation and beam analysis is important.

### 4. Properties of synchrotron radiation from steering magnets

The characteristics of the steering magnet (SM) magnetic field may differ significant from that of the BM. The magnetic field of the SM can be tuned over a wide range of values, and the length $L$ of magnetic field along the electron trajectory is much smaller than one of the BM. For instance for the SM of KEK ATF extraction line $L$=6 cm. Thus, the condition $R/\gamma << L$ for SM may not be fulfilled and the influence of both edges of magnetic field will take place. Nevertheless, our model allows us to calculate the SR distribution for these conditions. If $R/\gamma$ is comparable to $L$, then two parameters $\gamma$ and electron inflecting angle $\delta$ are important for the SR angular distribution. Fig.7 shows the calculated SR distribution from the SM of KEK ATF extraction line for different inflecting angles $\delta$. Analysys of these distributions shows that the SR differential intensity depends strongly on the inflecting angle $\delta$. For $\delta \approx 1$ mrad the intensity of SR from the SM may exceed that from the BM. Fig.8 shows the intensity in maximum of angular SR distribution as a function of electron inflecting angle.

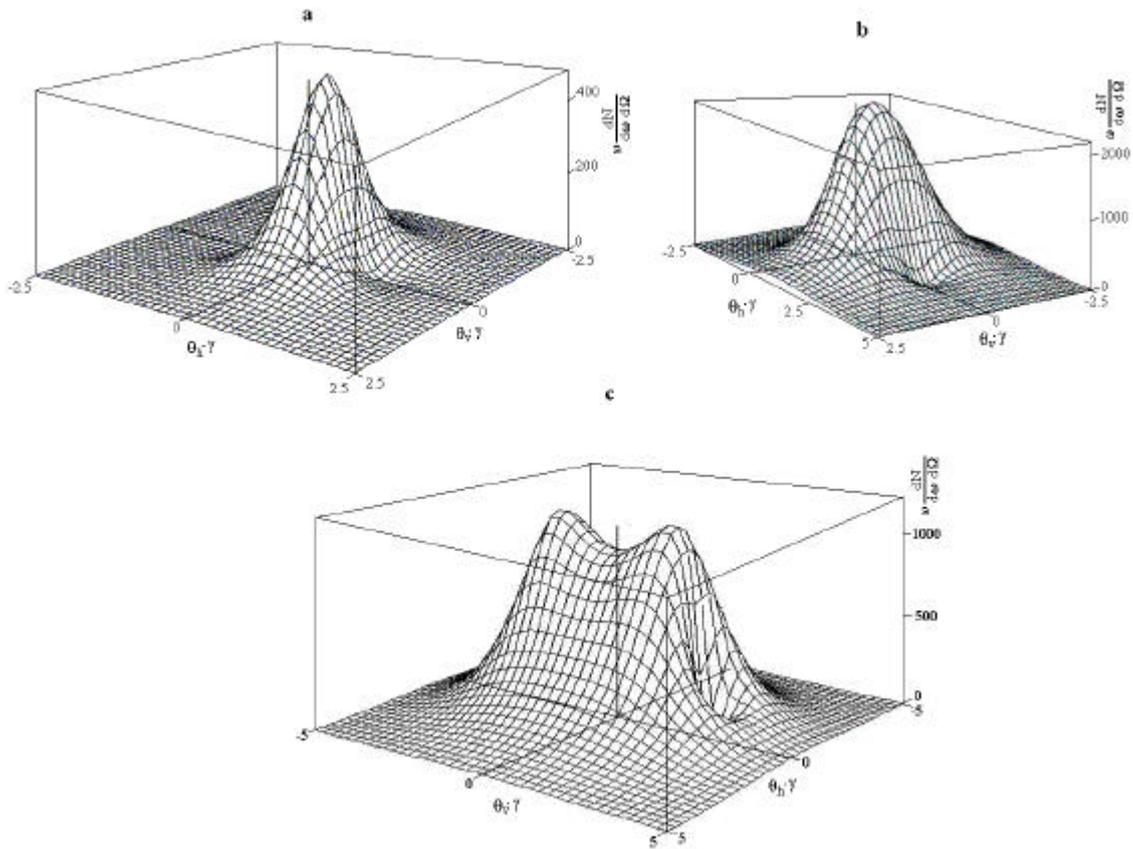

**Fig.7**

*The SR angular distribution from steering magnets for different inflecting angles d.*
*a) d=0.3 mrad,   b) d=1 mrad,   c) d=2 mrad,*

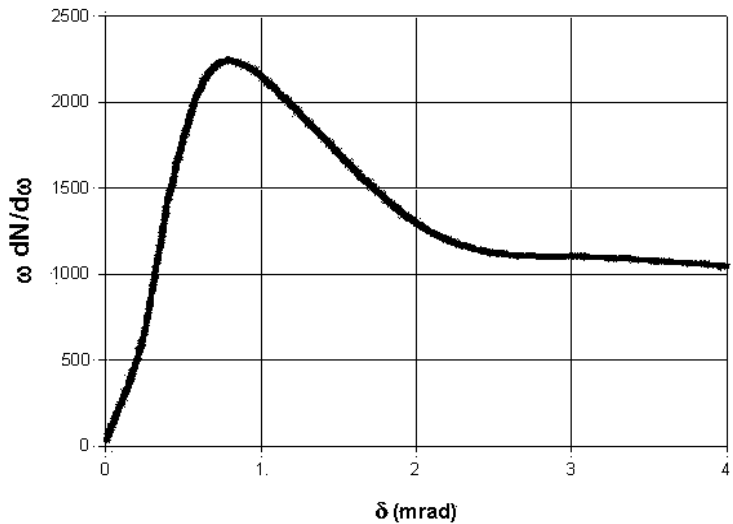

**Fig.8**

*The SR intensity from a SM for the maximum of angular distribution as a function of electron deflection angle.*

As can be seen, the SR from the steering magnets should be accounted for during BOTR investigations. The SR from the BM is usually stable while the SR from SM is difficult to predict because it depends strongly on the tuning and position of the electron beam.

## 5. Possible ways of SR suppression

Some laboratories [3,8] that use the BOTR for beam diagnostics on accelerators of similar experimental conditions have not experienced any problems from SR. However, the aim of their investigations was the observation of the BOTR spot on the target for beam size measurements. Usual microscopes of high quality are used for this purpose, but all information about BOTR angular distribution in this case is lost. The major characteristic of BODR for beam diagnostics is the angular distribution (see [1,2]); therefore we cannot directly use this method.

The analysis of the SR properties shows, that we can suppress significant the SR contribution by two ways: 1. As we can see on Fig.3, the yield of SR in the aperture size of $S$ in the center of angular distribution is proportional to $\approx S^4$. If we decrease the size of BODR target as much as possible, we decrease the contribution of the reflected SR. The limit of this possibility is determined by the possible violation of the semi-infinity plane approach. However, if the contribution of several magnets take place, this method may not be effective. 2.Another way may be realized using a bi-concave lens. We can use for this purpose the difference between the positions of the SR and BODR (or BOTR) souces. The principle of such suppression is shown on Fig.9.
We consider here the forward transition radiation (TR) and forward diffraction one (DR) for visual simplicity. The same principle can be used for backward radiation.
Even though the angular distributions that of the SR and DR (TR) are the same, the defocusing lens scatters the SR much more than that of the TR and the DR, as it is shown in the figure. Since the intensity of radiation is proportional to the $1/\theta^2$, the intensity of the SR after lens is attenuated more than that of the DR and the TR. The optics should be optimized to exclude the significant contribution in angular distribution of DR and TR source size on the target. The SR suppression using this method depends on the the distance from the magnet to the target, and from the target to the detector. It also depends on the DR (TR) source size on the target.

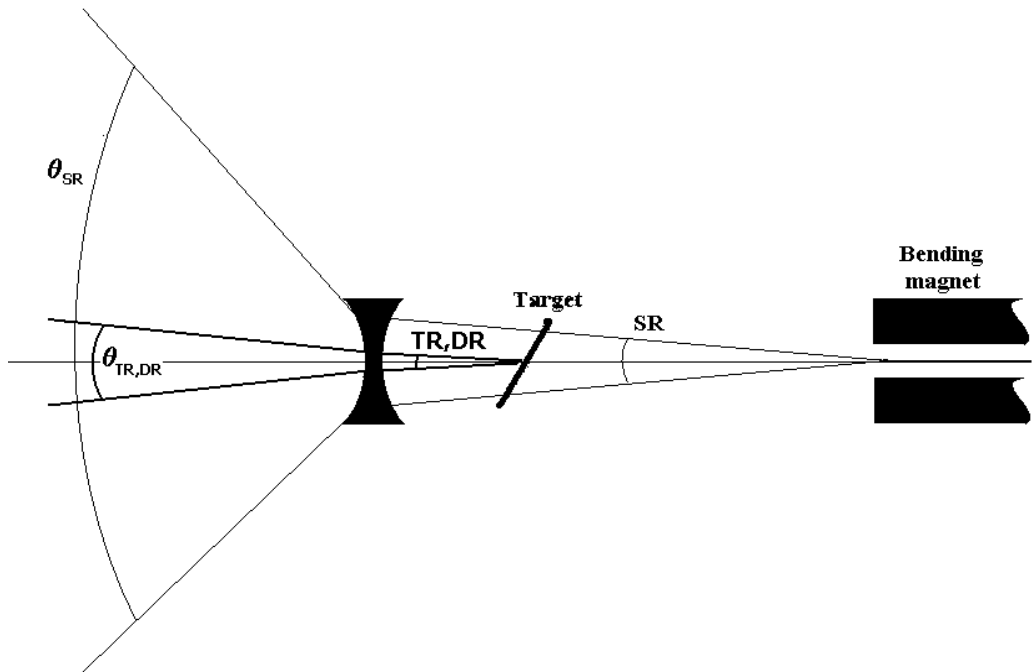

**Fig.9**

*A possible method of the optical SR suppression*

In conclusion, the problem of SR suppression may be difficult for determining angular BODR properties for the beam diagnostics and further attention is neeeded for its resolution. The size of BODR target should be decreased as much as possible. The SR suppression should be also forced using a bi-concave lens.

**Acknowledgements**

This work was supported by the RFBR (grant 99-02-16884) and by Program of state support of regional scientific high school politics of Russian Education Ministry and Atom-Ministry (grant 226).

**Referencies**